\documentclass[preprint,amsmath,amssymb]{revtex4}
\usepackage{graphicx}
\usepackage{dcolumn}
\usepackage{bm}
\usepackage{pstricks}
\usepackage{epsfig}
\usepackage{pst-grad}
\usepackage{pst-plot}
\begin{document}
\title{The Unruh Effect for Eccentric Uniformly Rotating Observers}
\author{${\rm H.\;Ramezani}$-${\rm Aval}$\footnote{Electronic
address:~ramezani@gonabad.ac.ir}} \affiliation{ Faculty of
Engineering Science, University of Gonabad, Gonabad, Iran}
\begin{abstract}
It is common to use Galilean rotational transformation to
investigate the Unruh effect for uniformly rotating observers.
However, the rotating observer in this subject is an eccentric
observer while Galilean rotational transformation is only valid for
centrally rotating observers. Thus, the reliability of the results
of applying Galilean rotational transformation to the study of the
Unruh effect might be considered as questionable. In this work the
rotational analog of the Unruh effect is investigated by employing
two relativistic rotational transformations corresponding to the
eccentric rotating observer, and it is shown that in both cases the
detector response function is non-zero. It is also shown that
although consecutive Lorentz transformations can not give a frame
within which the canonical construction can be carried out,
 the expectation value of particle number operator in canonical approach will be
zero if we use modified Franklin transformation. These conclusions
reinforce the claim that correspondence between vacuum states
defined via canonical field theory and a detector is broken for
rotating observers. Some previous conclusions are commented on and
some controversies are also discussed.
\end{abstract}
\maketitle
\section{Introduction} \label{sec1}
The Unruh effect predicts that linearly accelerated observer with
constant proper acceleration in flat spacetime (Rindler observer)
associates a thermal spectrum of particles to the no-particle state
(Minkowski vacuum) of inertial observer. In other words, two vacuum
states of these observers are not the same. So the particle content
of field theory is observer-dependent and observers with different
notions of positive and negative modes will disagree on the particle
content of a given state; what we think of as an inert vacuum
actually has the character of a thermal state. For a review on this
effect and its applications and experimental proposals see
\cite{Cris}. First approach to this effect was based on canonical
quantization of fields \cite{Fulling1,Davies1,Unruh1}, which we will
call "\emph{canonical approach}" and the second was based on
excitation of a detector \cite{Unruh1,Unruh2,DeWitt}, which we will
refer to it as "\emph{detector approach}". Despite the controversies
and disagreements in the interpretations and relations between them,
both approaches give the same mathematical result that an uniformly
accelerated detector (observer) observes a thermal spectrum of
particles and behaves as though it were placed in a thermal bath
with temperature $T=a/2\pi$, where $a$ is the magnitude of the
proper acceleration.

 A special case of accelerated observers which can have
more feasible experimental tests is eccentric uniformly rotating
observer. In addition to the usual ambiguities related to the
rotating observers \cite{Nouri1}, the particle detection due to
acceleration of rotational motion is also controversial and the
agreement between canonical approach and detector approach seems not
to occur for this observer. The problem of finding true rotational
transformation between the rotating and non-rotating frames is one
important aspect of conflict. While most authors have used Galilean
rotational transformation (GRT) to investigate the Unruh effect for
rotating observer
\cite{Denardo,Letaw,Davies,Letaw2,Grov,Suga,Ambrus}, some other have
tried to use corresponding relativistic transformations
\cite{Lore1,Lore2,Korsbakken,Faheem}. As we briefly mention them
below, their results are different .

 All of those who use GRT, in canonical approach, obtain zero
expectation value for particle number operator of rotating observer
in vacuum state of inertial observer, but they do not have agreement
on values (zero or nonzero) of detector response function
\cite{Birrel}. This problem was named \emph{the puzzle of rotating
detector} \cite{Lore1}. A solution for this problem was introduced
in \cite{Davies} which states that "\emph{confining
 the detector inside the limiting surface and imposing the speed of light restriction for detector, the rotating
 detector registers the absence of quanta and has vanishing response.}" Another explanation is that "\emph{the correspondence between expectation
 value for particle number operator defined via canonical quantum field theory and detector response function is  broken for
 general stationary motions}", and we must conclude that the two definitions are
 inequivalent \cite{Grov}. On the other hand, some of those who use relativistic transformations conclude the
 particle detection both for canonical and detector approaches\cite{Lore1,Lore2} and the other claim that there is no particle detection because rotating observe does not have
 event horizon\cite{Korsbakken}.

  In this paper we will discuss that GRT is not applicable for eccentric
  uniformly rotating observers and we must replace it with the correct relativistic transformations between laboratory inertial observer and
   eccentric rotating observer. We will use two sets of relativistic transformation to investigate unruh effect for eccentric uniformly rotating observer in
   canonical and detector approach. Also we will discuss other relativistic transformation mentioned above.

 In section II we briefly mention the limitations and problems of
 GRT and introduce two types of relativistic transformations for eccentric
 uniformly rotating observer that can be replaced with GRT. In section III we use two sets of relativistic transformations introduced in section II to investigate the Unruh effect
 for eccentric uniformly rotating observer both in canonical and detector approaches. We conclude with a
discussion section; comment on some former papers and discuss some
controversies.

 In this paper we use
$S^{'}$ for rest frame of laboratory observer and $S$ for
accelerated observer's frame. Greek letters take on the values
0,1,2,3. We use the metric with signature (+,-,-,-) and work in
natural units.
\section{Relativistic Transformations for Eccentric Uniformly Rotating Detector } \label{sec2}
As we have shown in \cite{Nouri1,Nouri2}, Galilean rotational
transformation (GRT)
\begin{eqnarray}\label{1}
t=t^\prime \;\;\; , \;\;\; r=r^\prime\;\;\; , \;\;\;
\phi=\phi^\prime- \Omega t  \;\;\; , \;\;\; z=z^\prime
\end{eqnarray}
for relation between centric inertial observer and eccentric
uniformly rotating observer who rotates with constant angular
velocity $\Omega$ at constant radius distance from the center of
rotation is not true; Specially absoluteness of time and not
distinguishing between observers at different radii in these
transformations cause inconsistent kinematical interpretations when
we want to explain phenomenon such as transverse Doppler effect and
Sagnac effect. This transformation is only applicable for relation
between two centric observer that one rotates uniformly and has no
translational motion and the other is a non-rotating inertial
observer. So using GRT for an eccentric rotating detector is not
true and the results that has been obtained by these transformations
for the Unruh effect in rotating frames are not valid.

 We assume a detector on the edge of a rigid disc that rotates uniformly
counterclockwise with angular velocity $\Omega$ in the $X^{'}Y^{'}$
plane around its axis ($Z^{'}$ axis). Such detectors are the ones
which are related to the real experimental setups. As we have shown
in \cite{Nouri2} there are two type of relativistic rotational
transformations to describe the relation between this rotating
observer (detector) and laboratory inertial observer:
\subsection{Special Relativistic Transformation (SRT)} SRTs are
based on consecutive Lorentz transformations and Fermi coordinates.
In \cite{MASH,NIKO} these coordinate transformations between
inertial laboratory (primed) and eccentric uniformly rotating
(unprimed) frames are given as follows
\begin{eqnarray}\label{2}
t=\gamma^{-1}(t^{\prime}-R\Omega\gamma
y)~~~x=x^{\prime}\sin(\gamma\Omega t)+y^{\prime}\cos(\gamma\Omega
t)-R \nonumber
\\
y=\gamma^{-1}[x^{\prime}\cos(\gamma\Omega
t)+y^{\prime}\sin(\gamma\Omega t)]~~~~,~~~~z=z^{\prime}
\end{eqnarray}
in which $\gamma=(1-R^2\Omega^2)^{-1/2}$, $\Omega$ is the uniform
angular velocity of the disk and R is the radius of the circular
path. In their setup the origin of the rotating frame is on the rim
of the circular path. The metric components in such a rotating frame
are given as follows
\begin{eqnarray}\label{3}
ds^2=-\gamma^{2}[1-(R+x)^{2}\Omega^2-\Omega^2 \gamma^{2}
{y}^{2}]{dt}^2+{dx}^2+{dy}^2+{dz}^2-2y\Omega{dxdt}+2x\Omega{dydt}
\end{eqnarray}
\subsection{Modified Franklin Transformations (MFT)}
In \cite{Nouri1}, looking for a consistent relativistic rotational
transformation between an inertial observer and an observer at a
non-zero radius (eccentric observer) on a uniformly rotating disk,
the following transformations (in cylindrical coordinates) were
introduced
\begin{eqnarray}\label{4}
t =  \cosh (\Omega R/c)t^{\prime} - \frac{R}{c}  \sinh (\Omega
R/c)\phi^{\prime} \;\;\; ; \;\;\; \rho = \rho^{\prime} \cr \phi =
\cosh (\Omega R/c) \phi^{\prime} -  \frac{c}{R} \sinh (\Omega R/c)
t^{\prime} \;\;\; ; \;\;\; z = z^{\prime},
\end{eqnarray}
in which $\Omega$ is the uniform angular velocity of the disk and
$R$ is the radial position of the observer on the disk. Note that
the origin of the rotating frame $S$ is chosen to be at the center
of the rotating disk so that both inertial and rotating frames
assign the same radial coordinate to the events. The corresponding
metric in the rotating observer's frame is given by,
\begin{eqnarray}\label{5}
ds^2 = -c^2\cosh^2(R\Omega)(1- \frac{{\rho}^2}{R^2}\tanh^2(R\Omega))
{dt}^2 + {d\rho}^2 + {\rho}^2 \cosh^2(R\Omega)  \cr (1-
\frac{R^2}{{\rho}^2}\tanh^2(R\Omega))d{\phi}^2 -2cR \sinh (R\Omega)
\cosh (R\Omega) (1- \frac{{\rho}^2}{R^2})dt d\phi  + {dz}^2.
\end{eqnarray}
\section{Particle Detection by Uniformly Rotating Eccentric Observer}\label{sec3}
Solving Klein-Gordon equation for a massless scalar field in flat
spacetime of laboratory inertial observer in cylindrical coordinate
gives positive modes solution as
\begin{eqnarray}\label{6}
f=\frac{1}{2\pi\sqrt{2\omega}}\exp(-i\omega
t'+im\phi'+ikz')J_m(q\rho')
\end{eqnarray}
 in which $m$ is an integer, $J_m$ is the Bessel function and $\omega=\sqrt{q^2+k^2}$. By expanding the field in term of a complete set of positive modes $f_i$
and negative modes ${f_i}^*$ and creation and annihilation
operators($\hat{a}_i^\dag$ and $\hat{a}_i$), we have
\begin{eqnarray}\label{7}
\Phi={\sum}_i{(\hat{a}_if_i+\hat{a}_{i}^{\dag}f^{*}_i)}
\end{eqnarray}
Also we can solve Klein-Gordon equation for a massless scalar field
in the rotating observer's frame and expand the field in term of a
new complete set of positive and negative modes $(g_i,g_i^{*})$ and
new creation and annihilation operators
$(\hat{b}_{i},\hat{b}_{i}^{\dag})$
\begin{eqnarray}\label{8}
\Phi={\sum}_i{(\hat{b}_ig_i+\hat{b}_{i}^{\dag}g^{*}_i)}.
\end{eqnarray}
If we show the vacuum state of inertial observer by $|0_f\rangle$
and the rotating observer's particle number operator by $\hat{n}_g$
then we have
\begin{eqnarray}\label{9}
\langle 0_f |\hat{n}_{gi}|0_f\rangle=\sum_j|\beta_{ij}|^2
\end{eqnarray}
in which the Bogolyubov coefficient $\beta$ is defined as
\begin{eqnarray}\label{10}
\beta_{ij}=-(g_i,{f_j}^*)
\end{eqnarray}
and definition of inner product is given by
\begin{eqnarray}\label{11}
(\varphi_1,\varphi_2)=-i\int_{\sum}(\varphi_1\nabla_\mu
\varphi_2^*-\varphi_2^*\nabla_\mu \varphi_1)n^\mu\sqrt{h}dx^{n-1}
\end{eqnarray}
where $\sum$ is hypersurface that we integrate over it, $h$ is the
determinant of  $h_{ij}$ which is the induced metric on $\sum$
 and $n^\mu$ is the normal vector to $\sum$ .
Since the inner product is independent of the hypersurface over
which the integral is taken, we can take the integral over $t=0$
hypersurface. According to (\ref{9}) non-zero value for coefficient
$\beta$ means non-zero expectation value of particle number operator
of the rotating observer in the vacuum state of laboratory observer. \\
On the other hand, according to \cite{Birrel} the detector response
function is given by
\begin{eqnarray}\label{12}
\emph{F}(E)=\int_{-\infty}^{\infty}d\triangle \tau
e^{-iE\triangle\tau}G^{+}(x(\tau_1),x(\tau_2))
\end{eqnarray}
where $G^{+}$ is the positive Wightman function defined as
\begin{eqnarray}\label{13}
G^{+}(x'_1,x'_2)=\frac{-1}{4\pi^2[(t'_1-t'_2-i\epsilon)^2-|\textbf{x}'_1-\textbf{x}'_2|^2]}
\end{eqnarray}
 and $x(\tau)$ is the
worldline of detector and $\tau$ is its proper time. \\
Now we calculate Bogolyubov coefficient $\beta$ and detector
response function using two relativistic rotational transformations
introduced in previous section.
\subsection{Special Relativistic Transformation (SRT)}
\subsubsection{Canonical Approach} By special relativistic
transformations (\ref{2}) and corresponding metric (\ref{3}) and
make two simplifying assumptions on the values of $R$ and $\Omega$
that $R=1$ and $\Omega=\frac{1}{2}$ (which are not important in our
discussion)we have
\begin{eqnarray}\label{14}
g_{00}=1-x^2/3-2x/3-4y^2/9~~,~~g_{01}=2y/3~~,~~g_{02}=-2x/3~~,g_{03}=0~~,~~g_{ij}=-\delta_{ij}
\end{eqnarray}
and so $g=-(x-3)^2/9$ and
\begin{eqnarray}\label{15}
g^{\mu\nu}=\frac{1}{(x-3)^2}\left(
  \begin{array}{cccc}
    9 & 6y & -6x & 0 \\
    6y & -x^2+6x+4y^2-9 & -4yx & 0 \\
    -6x & -4yx & 3(x^2+2x-3) & 0 \\
    0 & 0 & 0 & -1 \\
  \end{array}
\right)
\end{eqnarray}
and so the corresponding Klein-Gordon equation for massless scalar
field
($\frac{1}{\sqrt{-g}}\partial_\mu(\sqrt{-g}g^{\mu\nu}\partial_{\nu}\Phi)=0$)
is given by
\begin{eqnarray}\label{16}
\frac{9}{(x-3)^2}\frac{\partial^2\Phi}{\partial
t^2}+\frac{12y}{(x-3)^2}\frac{\partial^2\Phi}{\partial t \partial
x}-\frac{12x}{(x-3)^2}\frac{\partial^2\Phi}{\partial t \partial y}
 -\frac{6y}{(x-3)^3}\frac{\partial\Phi}{\partial
t}+\frac{4y^2-(x-3)^2}{(x-3)^2}\frac{\partial^2\Phi}{\partial
x^2}\nonumber \\
+\frac{12y^2-3(x-3)^2}{3(x-3)^3}\frac{\partial\Phi}{\partial
x}+\frac{3(x^2+2x-3)}{(x-3)^2}\frac{\partial^2\Phi}{\partial
y^2}=0~~~~~~~~~~~~~~~~~~~~~~~~~~~~~~~~~~~~~~~~~~
\end{eqnarray}
Although it seems necessary to obtain the analytic solution for this
partial differential equation to continue and calculate the
Bogolyubov coefficient, but it is not possible. As it mentioned very
shortly in \cite{Fulling2}, this is the reason why rotational
transformations based on consecutive Lorentz transformations can not
give a coordinate system within which the canonical approach of a
quantum field can be carried out.
\subsubsection{Detector Approach}
 With the assumption that the detector is at the origin of rotating frame and
using transformations (\ref{2}), the detector's trajectory in the
laboratory frame is given by
\begin{eqnarray}\label{21}
x'=R \cos(\gamma\Omega t)~~~,~~~y'=R \sin(\gamma\Omega
t)~~~,~~~z'=0~~~,~~~t'=\gamma t
\end{eqnarray}
 and by (\ref{13}) we have
\begin{eqnarray}\label{22}
G^{+}(x'_1,x'_2)=\frac{-1}{4\pi^2[(\gamma\triangle\tau-i\epsilon)^2-2R^2(1-\cos(\gamma\Omega
\triangle\tau))]}
\end{eqnarray}
Inserting (\ref{22}) in (\ref{12}) the detector response function is
given by
\begin{eqnarray}\label{23}
\emph{F}(E)=\int_{-\infty}^{\infty}d\triangle \tau
\frac{e^{-iE\triangle\tau}}{(\gamma\triangle\tau-i\epsilon)^2-4R^2\sin^2{(\gamma\Omega
\triangle\tau)}}.
\end{eqnarray}
Except in some constant coefficients this is the same as obtained
and numerically evaluated in \cite{Letaw} and so has non-zero
value.\\
\subsection{Modified Franklin Transformations (MFT)}
\subsubsection{Canonical Approach}
 For eccentric rotating observer using modified Franklin transformations (\ref{4}) and corresponding metric (\ref{5}) we
 have $g=-\rho^2$ and
\begin{eqnarray}\label{24}
g^{\mu\nu}=\left(
             \begin{array}{cccc}
               \cosh^2(R\Omega)-\frac{R^2\sinh^2(R\Omega)}{\rho^2} & 0 & \frac{R^2-\rho^2}{R\rho^2}\cosh(R\Omega)\sinh(R\Omega) & 0 \\
               0 & -1 & 0 & 0 \\
               \frac{R^2-\rho^2}{R\rho^2}\cosh(R\Omega)\sinh(R\Omega) & 0 & \frac{-\cosh^2(R\Omega)}{\rho^2}+\frac{\sinh^2(R\Omega)}{R^2} & 0 \\
               0 & 0 & 0 & -1 \\
             \end{array}
           \right)
\end{eqnarray}
 So the Klein Gordon equation is as bellow
\begin{eqnarray}\label{25}
(\cosh^2{\beta}-\frac{R^2}{\rho^2}\sinh^2{\beta})\frac{\partial^2\Phi}{\partial
t^2}-\frac{1}{\rho}\frac{\partial}{\partial\rho}(\rho\frac{\partial\Phi}{\partial\rho})-(\frac{\cosh^2{\beta}}{\rho^2}-\frac{\sinh^2{\beta}}{R^2})\frac{\partial^2\Phi}{\partial
\phi^2}\nonumber\\
+\frac{2(R^2-\rho^2)^2}{R\rho^2}\sinh{\beta}\cosh{\beta}\frac{\partial^2\Phi}{\partial
t\partial\phi}-\frac{\partial^2\Phi}{\partial
z^2}=0.~~~~~~~~~~~~~~~~~~~~~~~~~~~~~~~
\end{eqnarray}\label{26}
Assuming a trial solution as
\begin{eqnarray}
g=\exp(-i\omega't+im'\phi+ik'z)R(\rho)
\end{eqnarray}
and inserting in (\ref{25}), the radial part equation is
\begin{eqnarray}\label{27}
\frac{d^2R(\rho)}{d\rho^2}-\frac{1}{\rho}\frac{d
R(\rho)}{d\rho}+[(m'+\sqrt{2}\omega')^2+\frac{-\omega'^2+2m'^2-2\sqrt{2}\omega'm'}{\rho^2}]R(\rho)=0
\end{eqnarray}
in which we set $R=1$ and $\cosh^2(\beta)=2$ for simplicity. (These
assumptions do not affect the results.) This equation is a
cylindrical Bessel equation and so the positive mode solution
corresponding to it is
\begin{eqnarray}\label{28}
g=N\exp(-i\omega't+im'\phi+ik'z)J_m(q'\rho)
\end{eqnarray}
in which N is a normalization factor. As in the case of GRT
\cite{Denardo}, we can see that the Bogolyubov coefficient $\beta$
is zero here, so using MFT the canonical approach concludes the
absence of particle in the vacuum state of laboratory observer for the rotating observer.\\
\subsubsection{Detector Approach}
On the other hand the detector's trajectory in the rotating frame is
\begin{eqnarray}\label{29}
\rho=R~~,~~\phi=0~~,~~z=0
\end{eqnarray}
Using MFT to obtain the trajectory for the laboratory observer we
have
\begin{eqnarray}\label{30}
t'_2-t'_1=\cosh(R\Omega)\Delta\tau~~~,~~~\varphi'_2-\varphi'_1=\frac{\sinh(R\Omega)}{R}\Delta\tau
\end{eqnarray}
in which $\Delta\tau=t_2-t_1$. Wightman function in cylindrical
coordinate is as below
\begin{eqnarray}\label{31}
G^{+}(x'_1,x'_2)=\frac{-1}{4\pi^2}\frac{1}{(t'_2-t'_1-i\epsilon)^2-[{\rho'_2}^2+{\rho'_1}^2-2{\rho'_1}{\rho'_2}\cos(\varphi'_2-\varphi'_1)+(z'_2-z'_1)^2]}
\end{eqnarray}
 so we have
\begin{eqnarray}\label{32}
G^{+}(x'_1,x'_2)=\frac{-1}{4\pi^2[(\cosh(\beta)\triangle\tau-i\epsilon)^2-2R^2(1-\cos(\frac{\sinh(\beta)}{R}
\triangle\tau))]}
\end{eqnarray}
inserting in (\ref{12}) the detector response function is given by
\begin{eqnarray}\label{33}
\emph{F}(E)=\frac{1}{4\pi^2}\int_{-\infty}^{\infty}d\triangle \tau
\frac{e^{-iE\triangle\tau}}{(\cosh(\beta)\triangle\tau-i\epsilon)^2-4R^2\sin^2(\frac{\sinh(\beta)}{2R}
\triangle\tau)}
\end{eqnarray}
so in comparison with (\ref{23}) it has non-zero value.
\section{Discussion and Conclusions}
Here the Unruh effect for eccentric uniformly rotating observers was
investigated by two relativistic rotational transformations
corresponding to the eccentric rotating observer: consecutive
Lorentz transformation and modified Franklin transformation. It was
shown that the detector response function is non-zero in both cases.
We also showed that although consecutive Lorentz transformations
lead to calculational problem and give a frame within which the
Klein-Gordon equation does not have an analytic solution, but if we
use modified Franklin transformation, we obtain that the Bogolyubov
coefficient related to number operator and so the expectation value
of particle number operator is zero. This conclusions reinforce the
claim that correspondence between vacuum states defined via
canonical field theory and via a detector is broken for rotating
observers \cite{Letaw2,Grov}. Following our comparative study in
\cite{Nouri2}, here we showed that employing MFT instead of the SRT
helps to investigate the Unruh effect in canonical approach. It must
be emphasized that in these relativistic transformations the upper
limit for the velocity of the disk points (speed of light) is
 considered and unlike \cite{Davies} there is no need to confine the detector inside a light cylinder.
 In order to answer this question that if particle distribution is
characteristic of the thermal blackbody radiation with the finite
temperature, we need analytic solution of coefficient $\beta$ and
detector response function. Then we can judge the claim stated in
\cite{Korsbakken} that the Stationary detector will not show an
excitation spectrum which can be expressed simply in terms of
Boltzman factor.

 There are two important issues we face in
investigation of the unruh effect which seem to be the source of
this effect: the acceleration and the event horizon. Is the
existence of an event horizon a necessary condition for the Unruh
effect? What about acceleration? In \cite{Korsbakken} following up
\cite{Letaw2} the existence of horizon is assumed as a necessary
condition for creation of the Unruh effect; When there is an event
horizon we can define two different Fock space and mixing creation
and annihilation operators and will expect to have nonzero
Bogolyubov coefficients. Also they argue that "for a rotating
observer there is no event horizon since the orbit is restricted to
a bounded region of space, so that a signal
 from an event anywhere in space will be abe to reach the spacetime curve and any spacetime point can be reached by a light signal from a point on the
 curve." and conclude that for a uniformly rotating observer there
 is no corresponding unruh effect. The observer in our special relativistic approach which is the same as Mashhoon observer \cite{MASH}, has the $a/c<\omega$
 condition and is the same as uniformly rotating observer in \cite{Korsbakken} and so, according to it's result, should not
 observe Unruh effect. But as mentioned in \cite{Akhmedov} if the existence of horizon is necessary then even for linear accelerating detector the particle
 detection will be impossible, unless there is a detector with constant acceleration from the past infinity to future
infinity and this situations is practically inaccessible. On the
other hand if the acceleration is the necessary and sufficient
condition, then since the eccentric rotating observer has
centripetal acceleration, particle detection can be expected. The
remaining point is that the work done by the centripetal
  force in the case of uniform circular motion is zero and it can be an important differentiation between Rindler and uniformly rotating observer.

In \cite{Lore1,Lore2}using Trocheries-Takeno transformations, which
we called Franklin transformations (FT) \cite{Nouri1}, it is shown
that the rotating observer defines a vacuum state which is different
from the Minkowski one. But as we have discussed \cite{Nouri1}, FTs
have all kinematical problems of GRT and can not be applied for
relating eccentric rotating detector to centric laboratory observer.
In addition, in \cite{Lore2} the Klein-Gordon's solution that has
given in rotating frame is coordinate transformed solution of the
inertial one. But it is easy to show that by this assumption, unlike
their conclusion, always we will have zero Bogolyubov coefficient.
If $g(x')$ in (\ref{12}) is coordinate transformed of $f(x)$, when
we calculate integral (\ref{13}) we need to express $g$ and $f$ in
the same coordinates and need to apply inverse transformation on
$g$, so we will have
\begin{eqnarray}\label{34}
\beta=-(g(x'),{f}^*(x))=-(f(x),{f}^*(x))=i\int_{\sum}(f \nabla_\mu
f-f\nabla_\mu f)n^\mu\sqrt{-h}dx^{n-1}=0
\end{eqnarray}
and so it is impossible to obtain nonzero coefficient $\beta$ by
that suggested solution.\\
\textbf{Acknowledgments}\\
 The author thanks University of Gonabad for supporting
this project under the grants provided by the research council. Also
thanks Prof. M. Nouri-Zonoz for primary valuable suggestion.

\end{document}